\documentstyle[prd,preprint,tighten,aps,eqsecnum,amssymb,newlfont,epsfig]{revtex}
\begin{document}
\draft
\preprint{
\begin{tabular}{r}
   DFTT 46/98
\\ hep-ph/9808240
\end{tabular}
}
\title{A new ordering principle for the classical statistical analysis
of Poisson processes with background}
\author{C. Giunti}
\address{INFN, Sezione di Torino, and Dipartimento di Fisica Teorica,
Universit\`a di Torino,\\
Via P. Giuria 1, I--10125 Torino, Italy}
\maketitle
\begin{abstract}
Inspired by the recent proposal by
Feldman and Cousins \cite{FC98}
of a
``unified approach to the classical statistical analysis of small signals''
based on a choice of ordering
in Neyman's construction of classical confidence intervals,
I propose a new ordering principle
for the classical statistical analysis
of Poisson processes with background
which minimizes the effect
on the resulting confidence intervals
of the observation of less background events than expected.
The new ordering principle is applied to the calculation
of the confidence region implied by the recent null result of
the KARMEN neutrino oscillation experiment~\cite{KARMEN}.
\end{abstract}

\pacs{PACS numbers: 06.20.Dk, 14.60.Pq
\\
\\ Published in Phys. Rev. D \textbf{59}, 053001 (1999).
}

\section{Introduction}
\label{sec1}

In a very interesting paper \cite{FC98}
Feldman and Cousins
have recently proposed a
{``unified approach to the classical statistical analysis of small signals''}
that allows to construct
classical confidence belts which ``unify the treatment
of upper confidence limits for null results
and two-sided confidence intervals for non-null results''.
This unified approach is very attractive
because the transition from
two-sided intervals to upper limits
is automatic and the undercoverage introduced by
a choice based on the data
(``flip-flopping'')
is avoided by construction.
Indeed,
the unified approach has promptly been adopted by the
Particle Data Group
as the new standard method for statistical inference~\cite{PDG98}.

The probability to observe a number $n$
of events in a Poisson process
consisting in signal events with mean $\mu$
and background events with known calculated mean $b$
is
\begin{equation}
P(n|\mu,b)
=
\frac{1}{n!} \ (\mu+b)^n \, e^{-(\mu+b)}
\,.
\label{poisson}
\end{equation}
The classical method for obtaining the confidence interval
for the unknown parameter $\mu$
is based on Neyman's method to construct a \emph{confidence belt}.
This confidence belt is the region in the $\mu$-$n$ plane
lying between the two curves $n_1(\mu,\alpha)$ and $n_2(\mu,\alpha)$
such that for each value of $\mu$
\begin{equation}
P(n\in[n_1(\mu,\alpha),n_2(\mu,\alpha)]|\mu,b)
=
\alpha
\,,
\label{CL}
\end{equation}
where
\begin{equation}
P(n\in[n_1,n_2]|\mu,b)
=
\sum_{n=n_1}^{n_2} P(n|\mu,b)
\label{prob}
\end{equation}
and $\alpha$ is the desired confidence level.
The two curves
$n_1(\mu,\alpha)$ and $n_2(\mu,\alpha)$
are required to be monotonic functions of $\mu$
and can be inverted to yield the corresponding curves
$\mu_1(n,\alpha)$ and $\mu_2(n,\alpha)$.
Then,
if a number $n_{\mathrm{obs}}$ of events is measured,
the confidence interval for $\mu$ is
$[\mu_2(n_{\mathrm{obs}},\alpha),\mu_1(n_{\mathrm{obs}},\alpha)]$.
This method guarantees by construction the \emph{correct coverage},
\textit{i.e.}
the fact that the resulting confidence interval
$[\mu_2(n_{\mathrm{obs}},\alpha),\mu_1(n_{\mathrm{obs}},\alpha)]$
is a member of a set of confidence intervals
obtained with an ensemble of identical experiments
that
contain the true value of $\mu$ with a probability $\alpha$.
Actually,
in the case of a Poisson process,
since $n$ is an integer,
the relation (\ref{CL})
can only be approximately satisfied and in practice the chosen
\emph{acceptance intervals}
$[n_1(\mu,\alpha),n_2(\mu,\alpha)]$
are the smallest intervals such that
\begin{equation}
P(n\in[n_1(\mu,\alpha),n_2(\mu,\alpha)]|\mu,b)
\geq
\alpha
\,.
\label{CLP}
\end{equation}
This choice introduces an overcoverage for some values of $\mu$
and the resulting confidence intervals
are \emph{conservative}.
As emphasized in Ref.~\cite{FC98},
conservativeness is an undesirable but unavoidable property
of the confidence intervals in the case of a Poisson process
(it is undesirable because it
implies a loss of power
in restricting the allowed range for the parameter $\mu$).

The construction of Neyman's confidence belt
\emph{is not unique},
because in general there are many different couples of curves
$n_1(\mu,\alpha)$ and $n_2(\mu,\alpha)$
that satisfy the relation (\ref{CL}).
Hence,
an additional criterion is needed in order to
define uniquely the acceptance intervals
$[n_1(\mu,\alpha),n_2(\mu,\alpha)]$.
The two common choices are 
\begin{equation}
P(n<n_1(\mu,\alpha)|\mu,b)
=
P(n>n_2(\mu,\alpha)|\mu,b)
=
\frac{1-\alpha}{2}
\,,
\label{central}
\end{equation}
which leads to
\emph{central confidence intervals}
and
\begin{equation}
P(n<n_1(\mu,\alpha)|\mu,b)
=
1-\alpha
\,,
\label{upper}
\end{equation}
which leads to
\emph{upper confidence limits}.

Feldman and Cousins \cite{FC98}
proposed an ordering principle based on likelihood ratios
that produces an automatic transition
from a central confidence interval to an upper limit
when the number of observed events in a Poisson process with background
is of the same order or less than the expected background.
The acceptance interval for each value of $\mu$
is calculated assigning at each value of $n$ a rank
obtained from the relative size of the ratio
\begin{equation}
R(n)
=
\frac{ P(n|\mu,b) }{ P(n|\mu_{\mathrm{best}},b) }
\,,
\label{Rold}
\end{equation}
where $\mu_{\mathrm{best}}=\mu_{\mathrm{best}}(n)$
(for a fixed $b$)
is the non-negative value of $\mu$ that
maximizes the probability
$P(n|\mu,b)$:
\begin{equation}
\mu_{\mathrm{best}}(n)
=
\mathrm{max}[0,n-b]
\,.
\label{best}
\end{equation}
As emphasized in Ref.~\cite{FC98},
``$R$ is a ratio of two likelihoods:
the likelihood of obtaining $n$ given the actual mean $\mu$,
and the likelihood of obtaining $n$
given the best-fit physically allowed mean''.
The rank of each value of $n$
is assigned in order of decreasing value of the ratio $R(n)$:
the value of $n$ which has bigger $R(n)$ has rank one,
the value of $n$ among the remaining ones which has bigger $R(n)$ has rank two
and so on.
The acceptance interval for each value of $\mu$
is calculated by adding the values of $n$ in increasing order of rank
until the condition (\ref{CLP}) is satisfied.

It is clear that
the Feldman and Cousins ordering principle
guarantees an automatic transition from two-sided confidence intervals
to upper confidence limits for $ n \lesssim b $.
Indeed,
since
$ \mu_{\mathrm{best}}(n \leq b) = 0 $,
the rank of
$n \leq b$ for $\mu=0$
is one,
implying that the interval
$ 0 \leq n \leq b $
for $\mu=0$
is guaranteed to lie in the confidence belt.

Although the unified approach
solves brilliantly the problem of obtaining a
transition with correct coverage from two-sided confidence intervals
to upper confidence limits for $ n \lesssim b $,
it has the undesirable feature that even when
$ n \lesssim b $
the upper bound 
$\mu_1(n,\alpha)$
decreases when $b$ increases.
From a physical point of view
this is rather disturbing,
because a stringent upper bound for $\mu$
obtained by an experiment which has observed a number of events
significantly smaller than the expected background
is not due to the fact that the experiment is very sensitive to small values of $\mu$,
but to the fact that less background events than expected have been observed.
Hence,
two different experiments observing the \emph{same} Poisson process with mean $\mu$
and measuring the \emph{same} number of events
$n_{\mathrm{obs}}$
but expecting a different value for the background $b$
lead to two different upper limits for $\mu$
and the upper limit established by the experiment
that expect a larger background
can be significantly smaller than the other one.
This is illustrated in Figs.\ref{fig1} and \ref{fig2}.

In Fig.~\ref{fig1}
I have plotted the 90\% CL
confidence belts obtained with the unified approach
for $b=2,3,4,5$.
One can see that four experiments expecting, respectively,
$2$, $3$, $4$ and $5$ background events
and observing, for example,
$n_{\mathrm{obs}}=1$
lead to the 90\% CL
upper confidence limit
$\mu_1=2.53,1.88,1.34,1.20$,
respectively.

In Fig.~\ref{fig2}
I have plotted the upper end
$\mu_1$
of the 90\% CL
confidence intervals
obtained with the unified approach for $n=0,1,2,3,4,5$
as a function of the mean expected background $b$.
One can see that for each fixed value of $n$
the upper end
of the confidence intervals
tends to decrease rapidly
as the mean expected background $b$ increases
for small values of $b$
and stabilizes around a value close to 0.8
for large values of $b$.
The decrease of $\mu_1$ is not monotonic,
because of the discreteness of the observable $n$.
Figure \ref{fig2} differs from Fig.8 of Ref.~\cite{FC98}
for the fact that I have not forced $\mu_1$
to be a non-increasing function of $b$,
because,
as discussed above,
the decrease of the upper confidence limit when $b$ increases
is an undesirable feature from the physical point of view.
Therefore,
there is no reason to force $\mu_1$
to be a non-increasing function of $b$.
The only effect of this imposition is
to introduce an overcoverage
and then to increase the degree of conservativeness
of the upper confidence limits,
which is an undesirable effect.

The decrease of the upper confidence limit for a given $n$
as the mean expected background $b$ increases
is an unavoidable feature of the unified approach.
However,
I will show in the next Section that this decrease
can be weakened with the choice of
\emph{a new ordering principle}
for the construction of the confidence intervals.

\section{A new ordering principle}
\label{sec2}

The ordering principle introduced by Feldman and Cousins \cite{FC98}
is useful in order to guarantee an automatic transition from
two-sided confidence intervals to upper confidence limits.
However,
\emph{it is not unique}.
Therefore,
one can investigate the possibility to find a
new ordering principle which minimizes
the decrease of the upper confidence limit for a given $n$
as the mean expected background $b$ increases.
With this aim one can notice that the value (\ref{best}) of $\mu_{\mathrm{best}}(n)$
decreases proportionally to $n$ until it vanishes
an then it is forced to be non-negative ``by hand''.
It is clear that it would be preferable to have a reference value for $\mu$
which is automatically always positive.
Such a positive reference value
would imply that
for small but not vanishing values of $\mu$
the values of $n$ smaller than $b$
have a high rank.
As a consequence,
the decrease of the upper confidence limit for a given $n$
as the mean expected background $b$ increases
is weakened with respect to the one obtained with the
Feldman and Cousins ordering principle.

The new ordering principle
that I propose here
is based on the choice for the reference value for $\mu$
of the bayesian expectation value:
\begin{equation}
\mu_{\mathrm{ref}}(n)
=
\int_0^\infty \mu \, P(\mu|n,b) \, \mathrm{d}\mu
\,,
\label{mu_ref1}
\end{equation}
where $P(\mu|n,b)$
is the bayesian probability distribution for $\mu$
calculated assuming\footnote{This assumption is arbitrary,
but it seems to be the most reasonable one
if there is no prior information on the value of $\mu$,
which is the parameter probed directly in the experiment.}
a constant prior for $\mu\geq0$
(see, for example, \cite{D'Agostini}):
\begin{equation}
P(\mu|n,b)
=
( b + \mu )^n \, e^{-\mu}
\left( \displaystyle n! \, \sum_{k=0}^{n} \frac{b^k}{k!} \right)^{-1}
\,.
\label{bayes}
\end{equation}
The result of the integral (\ref{mu_ref1}) calculated by parts is
\begin{equation}
\mu_{\mathrm{ref}}(n)
=
n + 1
-
\left( \displaystyle \sum_{k=0}^{n} \frac{k\,b^k}{k!} \right)
\left( \displaystyle \sum_{k=0}^{n} \frac{b^k}{k!} \right)^{-1}
\,.
\label{mu_ref2}
\end{equation}
The obvious inequality
$
\sum_{k=0}^{n} k\,b^k/k!
\leq
n \sum_{k=0}^{n} b^k/k!
$
implies that
$ \mu_{\mathrm{ref}}(n) \geq 1 $.
Therefore,
$\mu_{\mathrm{ref}}(n)$
represents a reference value for $\mu$ that not only is non-negative,
as desired in order to have an automatic
transition from
two-sided intervals to upper limits,
but is even bigger or equal than one.
This is a desirable characteristics in order to
obtain a weak decrease of the upper confidence limit
for a given $n$
when the expected background $b$ increases.

The implementation of the new ordering principle
for the calculation of the confidence belt
is identical to the implementation of the Feldman and Cousins ordering principle,
except for the fact that in the calculation of the likelihood ratio $R(n)$,
which determines the rank of each value of $n$ for a fixed value of $\mu$,
$\mu_{\mathrm{best}}(n)$ is replaced by $\mu_{\mathrm{ref}}(n)$:
\begin{equation}
R(n)
=
\frac{ P(n|\mu,b) }{ P(n|\mu_{\mathrm{ref}},b) }
\,.
\label{Rnew}
\end{equation}
The resulting
90\% CL
confidence belts
for $b=2,3,4,5$
are shown in Fig.~\ref{fig3}.

Confronting Fig.~\ref{fig3} with Fig.~\ref{fig1}
one can see that the confidence belts are similar for $n \gg b$,
but the new upper confidence limits $\mu_1$ for
$ n \lesssim b $ are larger than the corresponding ones obtained
with the Feldman and Cousins ordering principle.
For example, four experiments expecting, respectively,
$2$, $3$, $4$ and $5$ background events
and observing
$n_{\mathrm{obs}}=1$
lead to the 90\% CL
upper confidence limits
$\mu_1=2.99,2.43,2.28,2.16$,
respectively,
which can be confronted with the respective values
$\mu_1=2.53,1.88,1.34,1.20$
obtained with the Feldman and Cousins ordering principle.
It is clear that the decrease of $\mu_1$
as $b$ is increased is milder with the new ordering principle.
This is fully illustrated in Fig.~\ref{fig4}
where I have plotted the upper end $\mu_1$
of the 90\% CL
confidence intervals
obtained with the new ordering principle for $n=0,1,2,3,4,5$
as a function of the mean expected background $b$.
One can see that for small values of $b$
the decrease of
the upper end $\mu_1$
of the confidence intervals
as a function of $b$
is weaker than in Fig.~\ref{fig2}
and
for large values of $b$ the value of
$\mu_1$
stabilizes around a value close to 1.7,
which is higher than in Fig.~\ref{fig2}.

Since the new ordering principle gives
upper bounds that are higher than
those obtained with the Feldman and Cousins ordering principle,
it must give also higher lower bounds.
This is confirmed by a comparison
of Figs.~\ref{fig1} and \ref{fig3}
and is illustrated in detail in Fig.~\ref{fig5},
where the values of the lower confidence limit $\mu_2$
obtained with the two methods
for $n=3,\ldots,9$ are plotted
as functions of the mean expected background $b$.
One can see that for each value of $n$ the maximum value of $b$
for which the lower confidence limit is different from zero
is the same in the two methods.
This is a consequence of the fact
(clearly seen with a comparison of Figs.~\ref{fig1} and \ref{fig3})
that the minimum value of $n$ for which there is a lower bound
different from zero
is the same in the two methods
(as must be,
since in both approaches
the lower limit of the acceptance interval for $\mu=0$
is at $n=0$).
Hence,
the experimental possibility to claim a positive effect
is the same in the two methods.
Furthermore,
Fig.~\ref{fig5} shows that the difference between the
lower limits obtained with the two methods
is moderate and
tends to vanish for $n \gg b$
(indeed,
for $n \gg b$
we have
$ \mu_{\mathrm{ref}} \simeq n = \mu_{\mathrm{best}} $
and the two methods are practically equivalent).

\section{Application to the result of the KARMEN experiment}
\label{sec3}

In this Section I discuss the application of the new ordering principle
proposed in Section \ref{sec2}
to the analysis of the recent null result of the KARMEN
neutrino oscillation experiment~\cite{KARMEN}.

The KARMEN experiment is searching for neutrino oscillations
(see, for example, \cite{neutrino oscillations})
in the $\bar\nu_\mu\to\bar\nu_e$
channel with a sensitivity in the region of the
neutrino mixing parameters $\sin^22\theta$ and $\Delta{m}^2$
which is allowed by the positive results of the LSND experiment~\cite{LSND}.
So far the KARMEN experiment
measured no events,
with an expected background of $ 2.88 \pm 0.13 $ events~\cite{KARMEN}.
In this case the unified approach leads to an upper confidence limit of 1.1 events
for the mean $\mu$ of neutrino oscillation events.
On the other hand,
the sensitivity of the experiment,
defined as
``the average upper limit one could get from an ensemble
of experiments with the expected background and no true signal'',
is of 4.4 events,
four times bigger than the upper confidence limit
(notice that the average upper limit is calculated using the unified approach).
This is clearly a pathological case
in which the observation of less background events than expected
leads to a stringent upper confidence limit $\mu_1$,
even if the experiment is not sensitive to the corresponding small
values of $\mu$.

The exclusion curves
(the confidence region lies on the left of the exclusion curves)
in the $\sin^22\theta$--$\Delta{m}^2$ plane
corresponding to the upper confidence limit of 1.1 events
obtained with the unified approach
and to the sensitivity of the experiment (4.4 events)
are reproduced in Fig.~\ref{fig6}
(the solid curves passing through the filled and empty circles, respectively).
The solid curve passing through the filled squares
corresponds to the upper limit of 2.3 signal events
obtained with the bayesian approach.
The shadowed area in Fig.~\ref{fig6}
is the region allowed at 90\% CL
by the results of the LSND experiment \cite{LSND}
and the dashed, dash-dotted and dash-dot-dotted curves are the
90\% CL
exclusion curves of the
Bugey \cite{Bugey}, BNL E776 \cite{BNL E776} and CCFR \cite{CCFR} experiments,
respectively.

The stringent limit for the neutrino mixing parameters
represented by the unified approach exclusion curve
strongly depends on the fact that
none of the expected background events has been observed.
Indeed,
the unified approach exclusion curve
is based on the 90\% CL
exclusion of a
mean signal of 1.1 events,
but the signal that would have been excluded
if KARMEN had observed, for example,
two (background) events
would have been 3.1 events
and the corresponding exclusion curve
in the
$\sin^22\theta$--$\Delta{m}^2$
plane would lie on the right of the bayesian exclusion curve.
In other words,
the fact that the unified approach exclusion curve
is very stringent
is not due to the fact that no neutrino oscillations were observed,
but to the fact that no background event
has been observed.
This is clearly undesirable from the physical point of view.

The new ordering principle proposed in the previous Section
allows to obtain an exclusion curve
with the correct coverage
that minimizes the effect of the fact
that no background event
has been observed in the KARMEN experiment.
The new ordering principle gives
a 90\% CL
upper confidence limit of 1.9 events
for the mean $\mu$ of neutrino oscillation events.
This limit is close to the upper limit of
2.3 events
obtained in the bayesian approach.
As shown in Fig.~\ref{fig6},
the exclusion curve corresponding to the upper confidence limit
obtained with the new ordering principle
(the solid curve passing through the filled triangle)
lies close to the bayesian exclusion curve.

The sensitivity of the KARMEN experiment calculated with the new ordering principle is
of 4.7 events,
rather close to the sensitivity of 4.4 events calculated with the unified approach.
The corresponding curve is shown in Fig.~\ref{fig6}
(the solid curve passing through the empty triangles).

Looking at Fig.~\ref{fig6}
one can see that the new ordering principle
allows to obtain from the results of the KARMEN experiment
an exclusion curve that is more reliable than the one
obtained with the unified approach,
because the discrepancy between the exclusion curve and the corresponding sensitivity
curve is smaller.
Furthermore,
since the new ordering principle implies an exclusion curve close to that
obtained with the bayesian approach,
we have a nice and desirable agreement of the results obtained with different
statistical methods,
one of which
(the new ordering principle)
guarantees a correct coverage.

In conclusion of this Section,
I would like to emphasize that,
in spite of the improvement
for the physical interpretation of the result of the KARMEN experiment
obtained with the introduction of the new ordering principle,
the fact that the KARMEN exclusion curves obtained 
with different statistical methods are significantly different
shows that no firm physical conclusion can be inferred
from the null result of the KARMEN experiment.
Since the KARMEN experiment is continuing to take data,
some background events consistent with the expected rate
should be observed in the near future
and
the different curves obtained with different statistical methods
should converge.
Only in that case it will be possible to confirm or exclude
the LSND indication in favor of neutrino oscillations.

\section{Conclusions}
\label{sec4}

The new ordering principle
for the classical statistical analysis
of Poisson processes with background
proposed here
allows to improve the unified approach
recently proposed by Feldman and Cousins~\cite{FC98}.

The confidence intervals obtained with the new ordering principle
take advantage of the desirable features of the unified approach
and
of the bayesian approach.
The choice of the bayesian expectation value (\ref{mu_ref1})
as the reference value $\mu_{\mathrm{ref}}(n)$
for the new ordering
guarantees a reference value bigger or equal than one.
The implementation of the new ordering principle
in the same way as
the Feldman and Cousins ordering principle
is implemented in the unified approach
guarantees that the resulting confidence intervals have the correct coverage
and that
there is an automatic transition from two-sided
intervals to upper limits for $n \lesssim b$.
The fact that the reference value
$\mu_{\mathrm{ref}}(n)$
is $\geq1$
implies that the decrease of the upper confidence limit $\mu_1$
for the mean signal $\mu$
as the mean expected background $b$
increases is weaker with the new ordering principle
than with the Feldman and Cousins ordering principle.
This is a desirable feature from a physical point of view,
because it allows to minimize the influence that the observation
of less background events than expected
has on the upper confidence limit
for the mean true signal events.

If one does not like the fact that
in the new ordering principle
the reference value
$\mu_{\mathrm{ref}}(n)$
has been obtained using the bayesian method,
one can consider Eq.(\ref{mu_ref2})
as the formula that allows to define
$\mu_{\mathrm{ref}}(n)$
in order to obtain the desired result
(as the conditions (\ref{central}) and (\ref{upper})
allow to obtain central confidence intervals and
upper confidence limits,
respectively).
After all, the construction of the confidence belt,
including the ordering principle,
is not uniquely defined from first principles
and a scientist can choose the method
which maximizes the scientific meaning
of the resulting confidence intervals,
as long as the correct coverage is guaranteed.

In Section~\ref{sec3},
I have applied the new ordering principle
to the analysis of the null result of the
KARMEN neutrino oscillation experiment~\cite{KARMEN}.
The new ordering principle
gives an exclusion curve
in the plane of the neutrino mixing parameters
$\sin^22\theta$ and $\Delta{m}^2$
which lies close to that obtained with the bayesian approach
and is significantly less stringent
than the one obtained with the unified approach.
Since the stringency of the
unified approach exclusion curve
is due to the fact that no background events have been observed
(with 2.88 mean expected background events)
and not to the sensitivity of the experiment
to
a small true neutrino oscillation signal,
I think that
the exclusion curve obtained with the new ordering
represents more appropriately
the physical implications
of the null result of the KARMEN experiment.

Finally,
I would like to notice that the new ordering principle has been
formulated here for the case of a Poisson process with background,
but can be generalized
to other controversial cases,
as that of gaussian errors
with a bounded physical region~\cite{in preparation}. 

\acknowledgments

I would like to thank K. Eitel
for sending me useful information on the statistical analysis
performed by the KARMEN collaboration
and C.W. Kim for useful discussions.
I would like to express my gratitude to the Korea Institute for Advanced Study
(KIAS)
for the kind hospitality during the initial stages of this work.

\begin{figure}[h]
\refstepcounter{figure}
\label{fig1}
Fig.~\ref{fig1}.
Confidence belts for 90\% CL
obtained with the unified approach
for a Poisson process with background
$b=2$ (region between the two solid lines),
$b=3$ (region between the two dashed lines),
$b=4$ (region between the two dash-dotted lines),
$b=5$ (region between the two dash-dot-dotted lines).
\end{figure}

\begin{figure}[h]
\refstepcounter{figure}
\label{fig2}
Fig.~\ref{fig2}.
Upper end of the 90\% CL
confidence intervals
obtained with the unified approach
for $n=0,1,2,3,4,5$
as a function of the mean expected background $b$.
\end{figure}

\begin{figure}[h]
\refstepcounter{figure}
\label{fig3}
Fig.~\ref{fig3}.
Confidence belts for 90\% CL
obtained with the new ordering principle
for a Poisson process with background
$b=2$ (region between the two solid lines),
$b=3$ (region between the two dashed lines),
$b=4$ (region between the two dash-dotted lines),
$b=5$ (region between the two dash-dot-dotted lines).
\end{figure}

\begin{figure}[h]
\refstepcounter{figure}
\label{fig4}
Fig.~\ref{fig4}.
Upper end of the 90\% CL
confidence intervals
obtained with the new ordering principle
for $n=0,1,2,3,4,5$
as a function of the mean expected background $b$.
\end{figure}

\begin{figure}[h]
\refstepcounter{figure}
\label{fig5}
Fig.~\ref{fig5}.
Lower end of the 90\% CL
confidence intervals
obtained with the unified approach and
with the new ordering principle
for $n=3,\ldots,9$
(the solid line, \ldots, the dash-dot-dotted line)
as a function of the mean expected background $b$.
\end{figure}

\begin{figure}[h]
\refstepcounter{figure}
\label{fig6}
Fig.~\ref{fig6}.
90\% CL
exclusion curves
in the plane of the neutrino oscillation parameters
$\sin^22\theta$--$\Delta{m}^2$
corresponding to the
null result of the KARMEN experiment~\cite{KARMEN}.
The solid curves passing through the filled squares, circles and triangles
are obtained with the bayesian approach, the unified approach
and the new ordering principle, respectively.
The solid curves passing through the empty circles and triangles
are the sensitivity curves obtained with the unified approach
and the new ordering principle, respectively.
The shadowed area
is the region allowed at 90\% CL
by the results of the LSND experiment \cite{LSND}
and the dashed, dash-dotted and dash-dot-dotted curves are the
90\% CL
exclusion curves of the
Bugey \cite{Bugey}, BNL E776 \cite{BNL E776} and CCFR \cite{CCFR} experiments,
respectively.

\end{figure}


\newpage

\begin{minipage}[p]{0.95\linewidth}
\begin{center}
\mbox{\epsfig{file=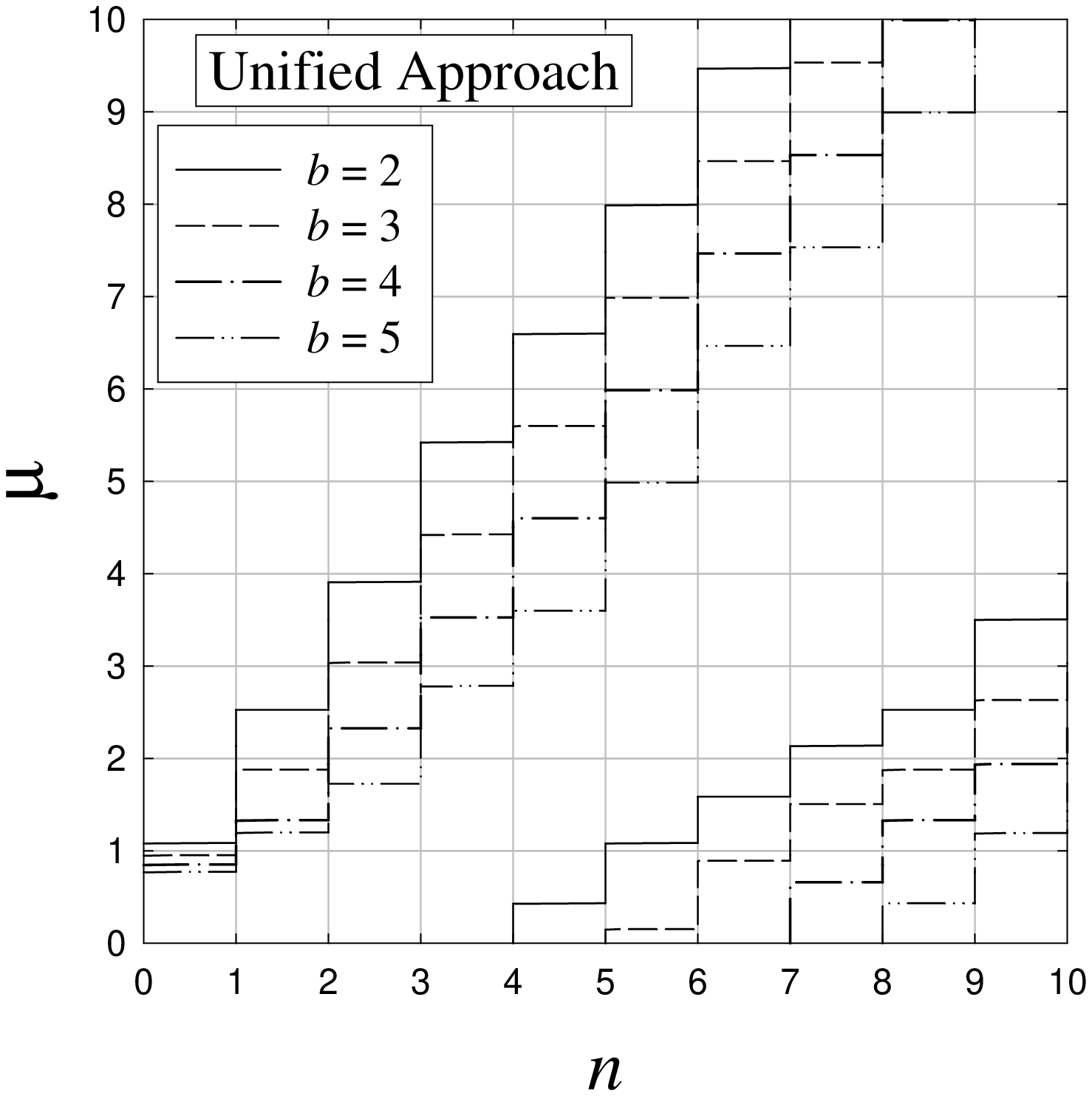,width=0.95\linewidth}}
\end{center}
\end{minipage}
\begin{center}
\Large Figure \ref{fig1}
\end{center}

\begin{minipage}[p]{0.95\linewidth}
\begin{center}
\mbox{\epsfig{file=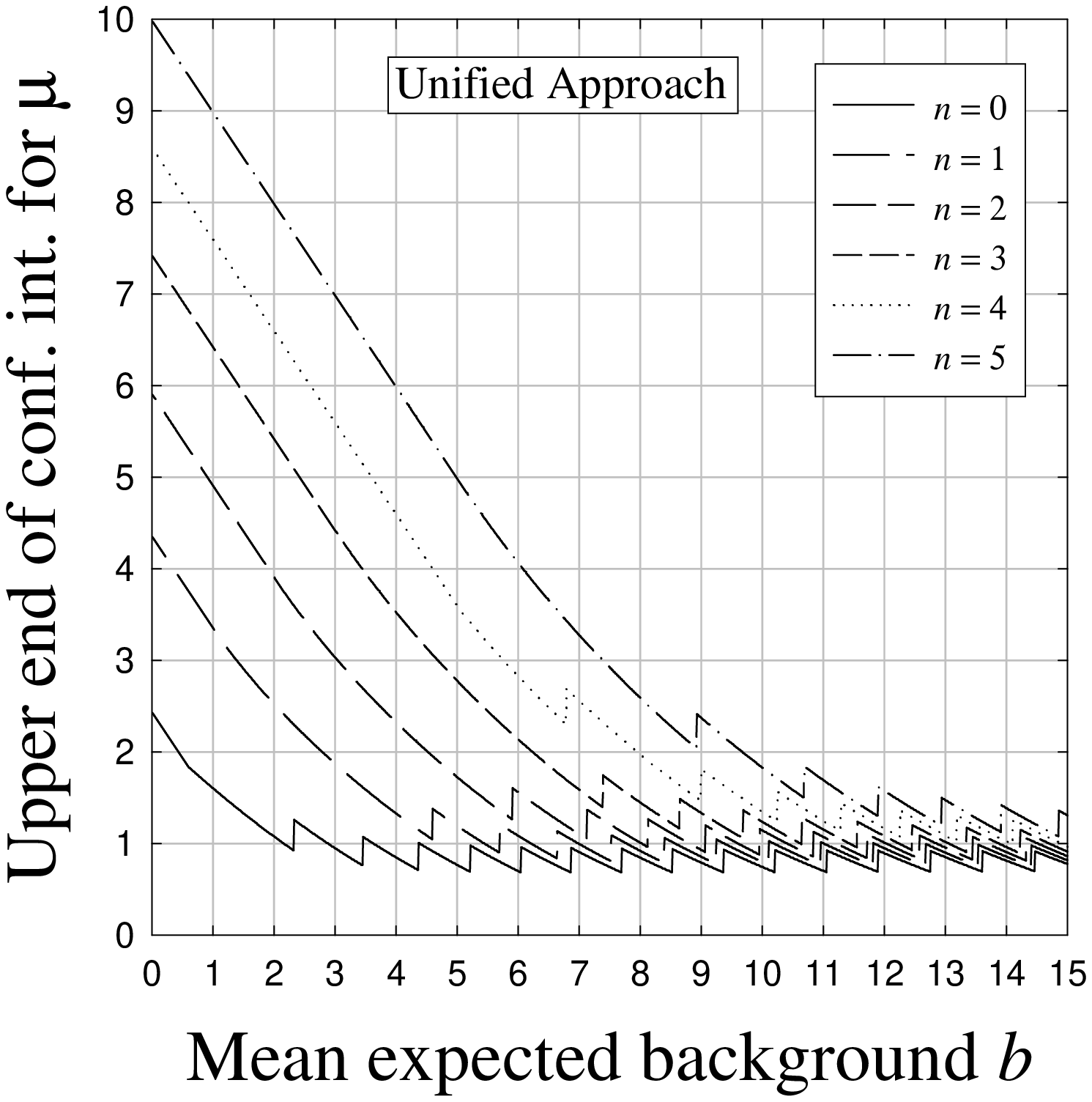,width=0.95\linewidth}}
\end{center}
\end{minipage}
\begin{center}
\Large Figure \ref{fig2}
\end{center}

\begin{minipage}[p]{0.95\linewidth}
\begin{center}
\mbox{\epsfig{file=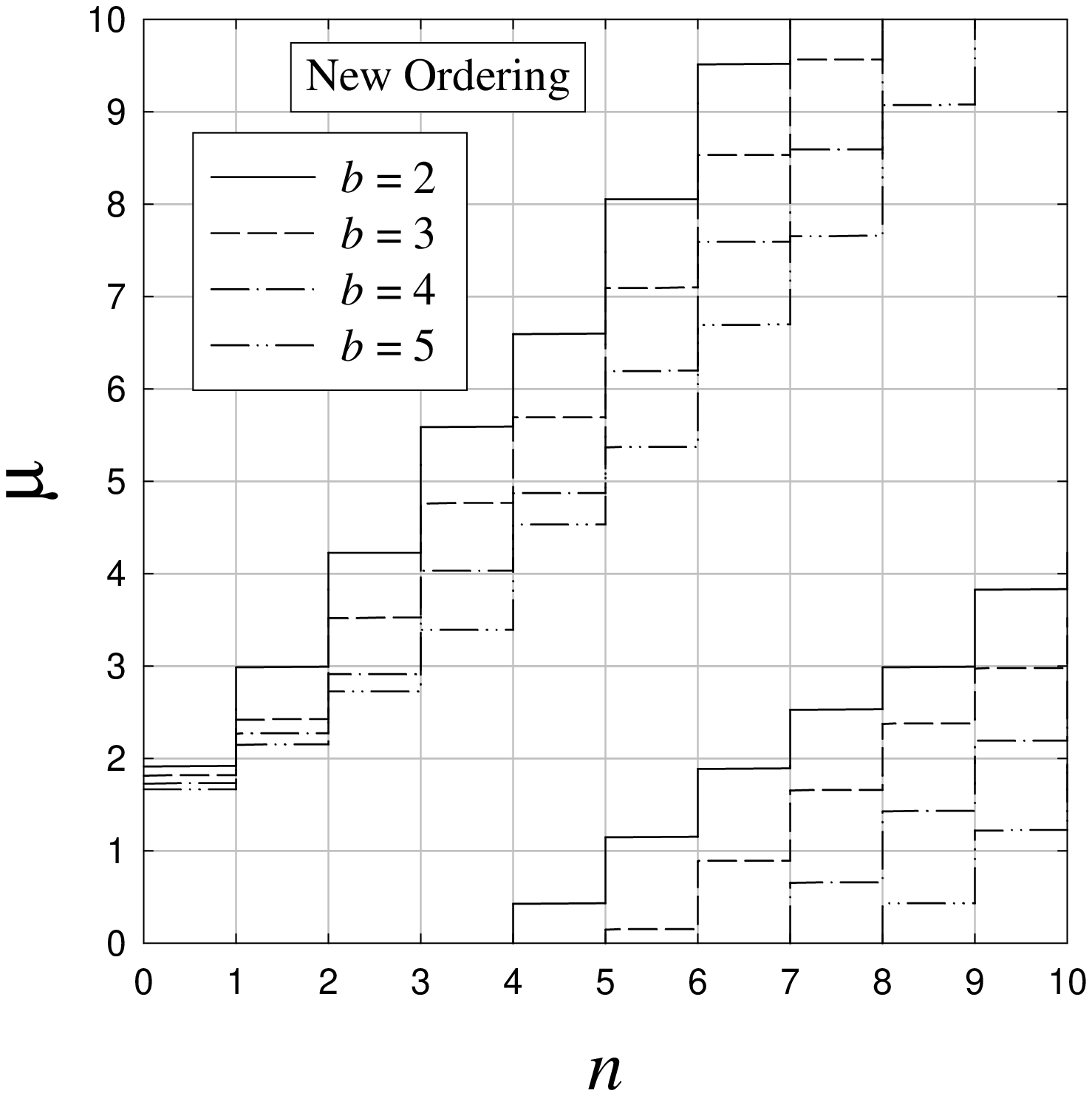,width=0.95\linewidth}}
\end{center}
\end{minipage}
\begin{center}
\Large Figure \ref{fig3}
\end{center}

\begin{minipage}[p]{0.95\linewidth}
\begin{center}
\mbox{\epsfig{file=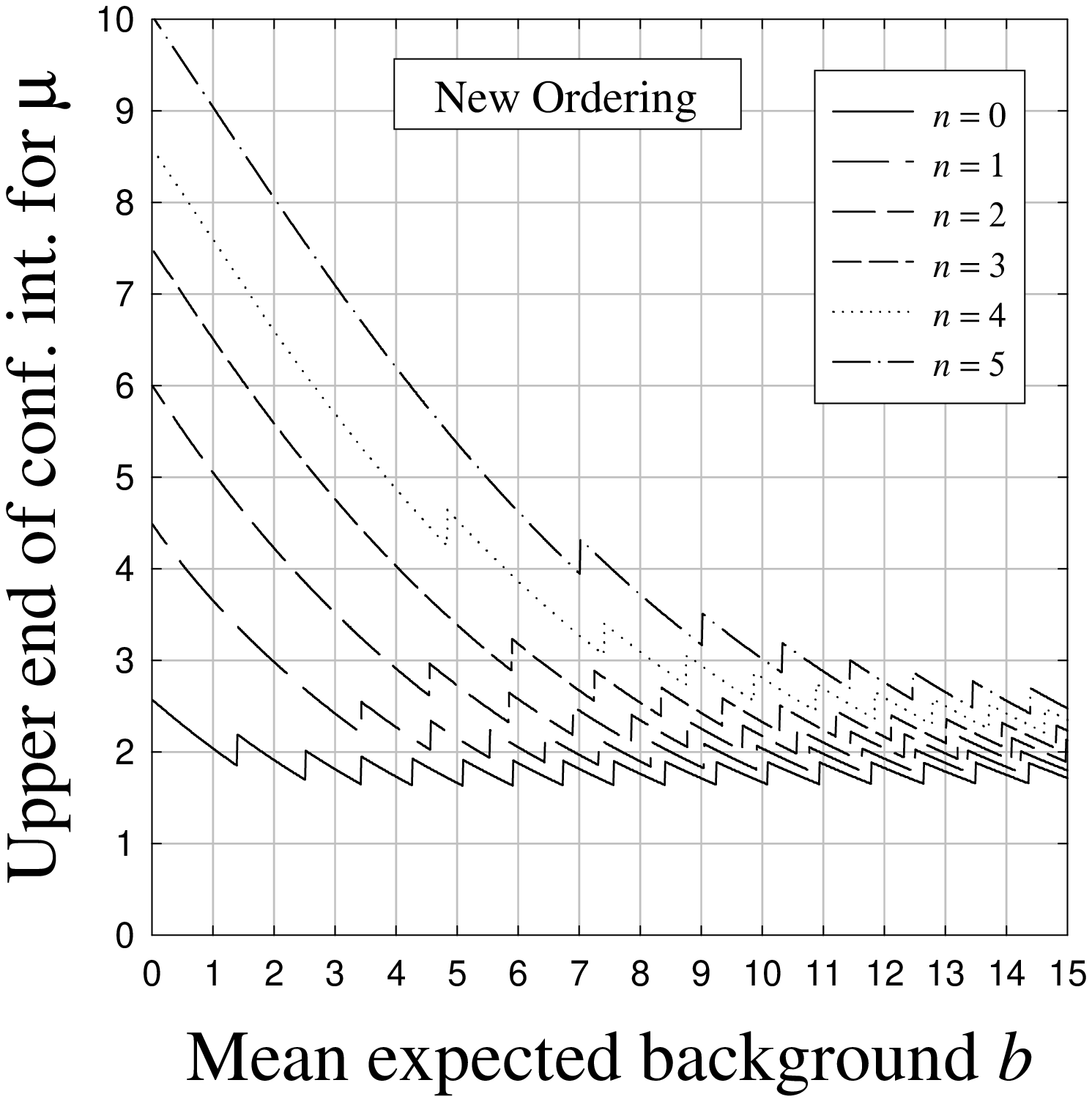,width=0.95\linewidth}}
\end{center}
\end{minipage}
\begin{center}
\Large Figure \ref{fig4}
\end{center}

\begin{minipage}[p]{0.95\linewidth}
\begin{center}
\mbox{\epsfig{file=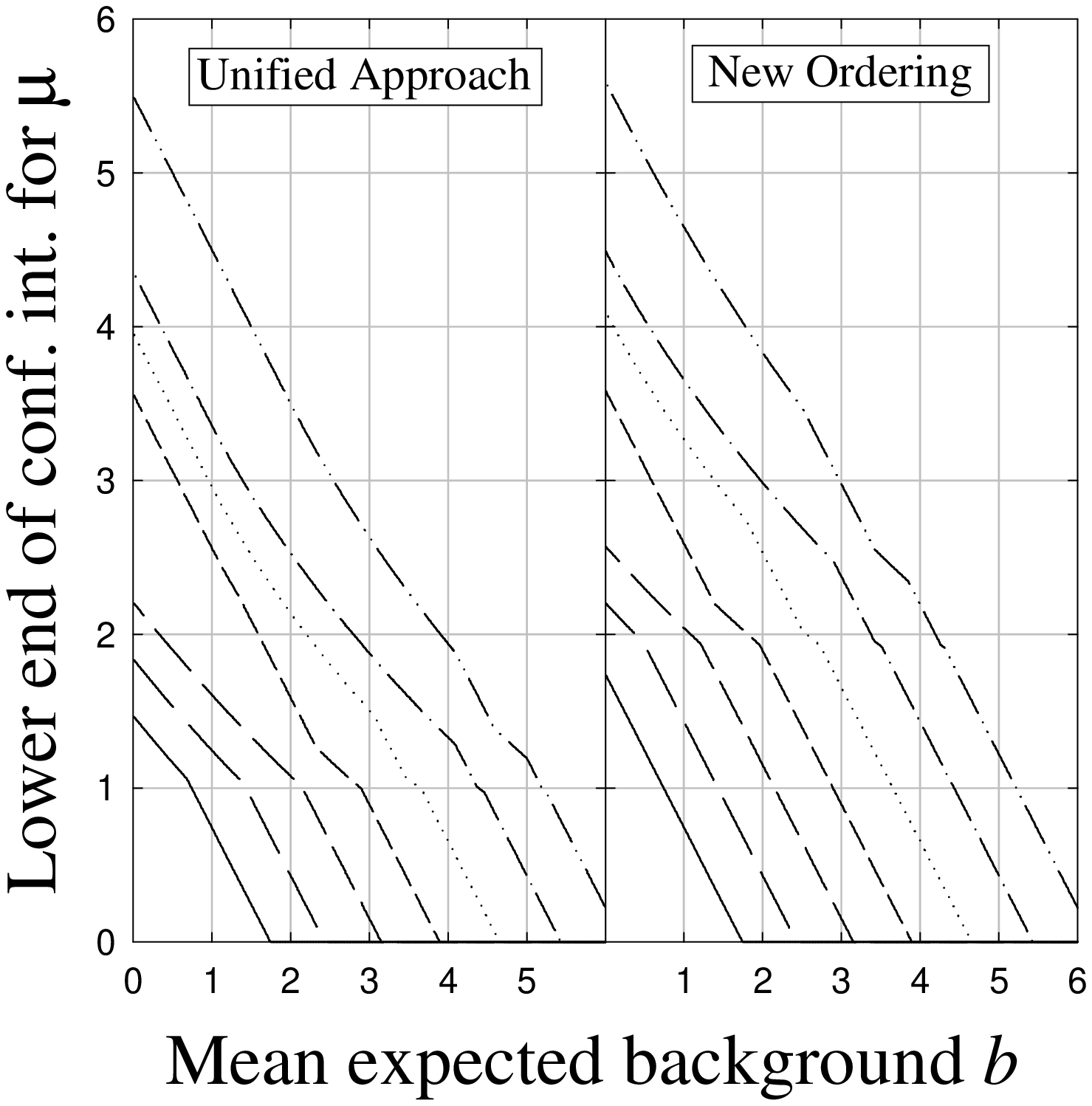,width=0.95\linewidth}}
\end{center}
\end{minipage}
\begin{center}
\Large Figure \ref{fig5}
\end{center}

\begin{minipage}[p]{0.95\linewidth}
\begin{center}
\mbox{\epsfig{file=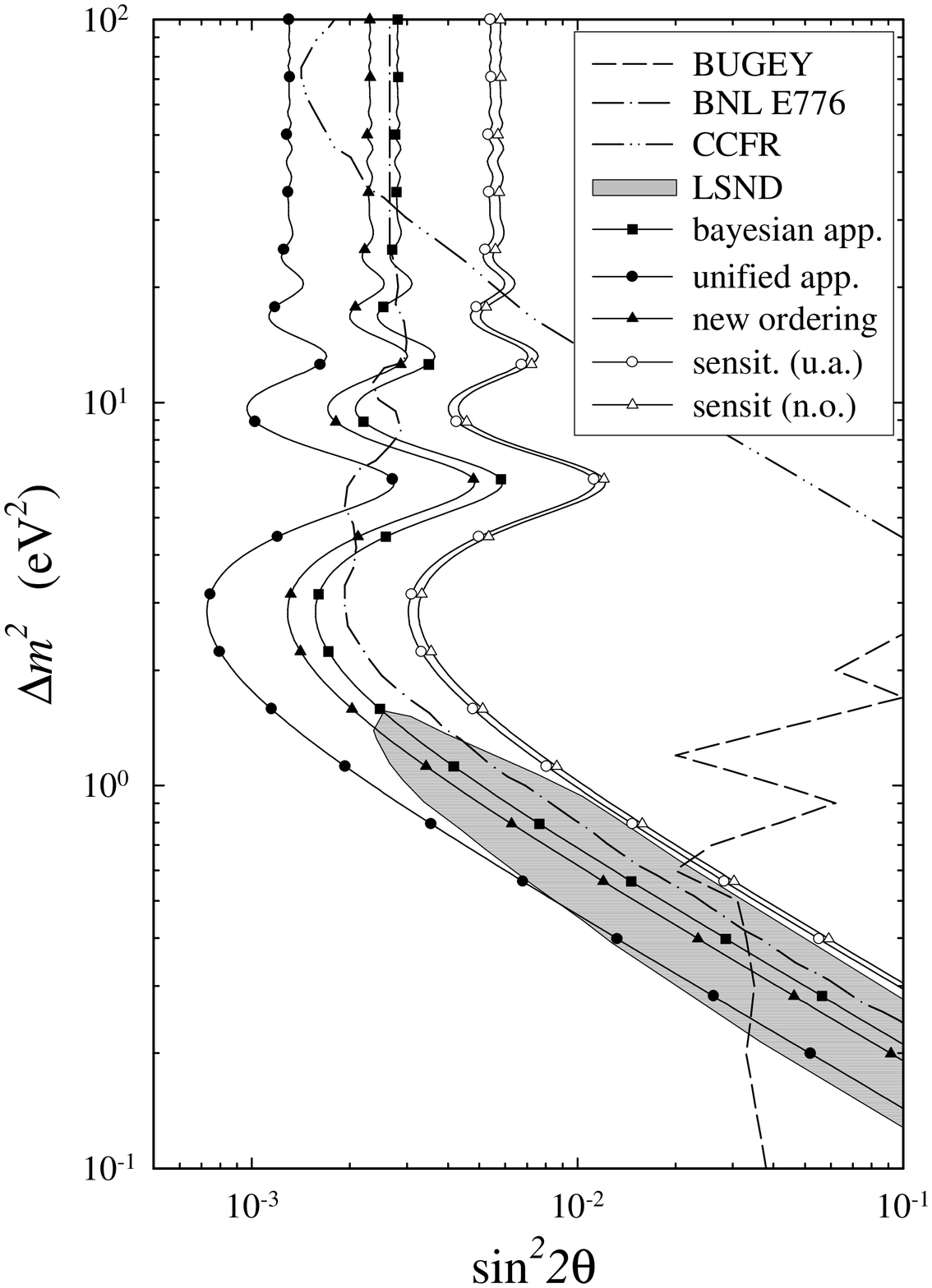,width=0.95\linewidth}}
\end{center}
\end{minipage}
\begin{center}
\Large Figure \ref{fig6}
\end{center}


\begin{references}

\bibitem{FC98}
G.J. Feldman and R.D. Cousins,
Phys. Rev. D \textbf{57}, 3873 (1998).

\bibitem{KARMEN}
B. Zeitnitz,
Talk presented at
\textit{Neutrino '98}
\cite{nu98};
KARMEN WWW page:
http://\-www-\-ik1.\-fzk.\-de/\-www/\-karmen/\-karmen\_e.\-html.

\bibitem{PDG98}
C. Caso et al,
Eur. Phys. J. C \textbf{3}, 1 (1998).

\bibitem{D'Agostini}
G. D'Agostini,
preprint DESY 95-242 (hep-ph/9512295).

\bibitem{neutrino oscillations}
S.M. Bilenky and B. Pontecorvo,
Phys. Rep. \textbf{41}, 225 (1978);
S.M. Bilenky and S.T. Petcov,
Rev. Mod. Phys. \textbf{59}, 671 (1987);
R.N. Mohapatra and P.B. Pal,
\textit{Massive Neutrinos in Physics and
Astrophysics},
World Scientific Lecture Notes in Physics, Vol.41
(World Scientific, Singapore, 1991);
C.W. Kim and A. Pevsner,
\textit{Neutrinos in Physics and Astrophysics},
Contemporary Concepts in Physics, Vol.8
(Harwood Academic Press, Chur, Switzerland, 1993).

\bibitem{LSND}
C. Athanassopoulos \textit{et al.},
Phys. Rev. Lett. \textbf{77}, 3082 (1996);
D.H. White,
Talk presented at
\textit{Neutrino '98}
\cite{nu98}.

\bibitem{Bugey}
B. Achkar \emph{et al.},
Nucl. Phys. B \textbf{434}, 503 (1995).

\bibitem{BNL E776}
L. Borodovsky et al.,
Phys. Rev. Lett. \textbf{68}, 274 (1992).

\bibitem{CCFR}
A. Romosan \emph{et al.},
Phys. Rev. Lett. \textbf{78}, 2912 (1997).

\bibitem{in preparation}
C. Giunti,
in preparation.

\bibitem{nu98}
\textit{Neutrino '98} WWW page:
http://\-www-\-sk.\-icrr.\-u-\-tokyo.\-ac.\-jp/\-nu98.

\end{references}
\end{document}